# Metallic ferroelectric-ferromagnetic multiferroics in strained EuTiO$_{3-x}$H$_x$

Sheng Xu[1], Yanni Gu[1,2], Xiao Shen[3*]

[1]School of Metallurgy and Materials Engineering, Jiangsu University of Science and Technology, Zhangjiagang, Jiangsu 215600, China

[2]National Laboratory of Solid State Microstructures and Department of Physics, Nanjing University, Nanjing, 210093, China

[3]Department of Physics and Materials Science, University of Memphis, Memphis, Tennessee 38152, USA

**Abstract:**

Polar metals are defined by the coexistence of metallicity and polar crystal structure. They have potential applications in non-linear optics, ferroelectric devices, and quantum devices. Meanwhile, ferroelectric-ferromagnetic (FE-FM) multiferroics display simultaneous ferroelectricity and ferromagnetism, leading to new technologies in information storage. It remains an open question whether metallicity, ferroelectricity, and ferromagnetism can coexist in a single domain of a material. EuTiO$_3$ is actively studied for potential applications in magnetic sensors, memories, magneto-optical devices, and energy conversion devices. It stabilizes at a multi-critical equilibrium and exhibits a rich range of intriguing properties. Here, using the results from hybrid density functional theory calculations, we report metallic FE-FM multiferroics in strain-engineered epitaxial EuTiO$_3$ with H doping. The emergence of the magnetism in polar metals provides a new degree of freedom to control these materials in applications. The underlying mechanism for the coexistence of metallicity, ferroelectricity, and ferromagnetism is discussed. The ferromagnetism in metallic EuTiO$_{3-x}$H$_x$ is explained by the Ruderman-Kittel-Kasuya-Yosida (RKKY) interaction, which agrees with experiments. The coexistence of metallicity and ferroelectricity is allowed because the electrons at the Fermi level are weakly coupled to the ferroelectric distortion. Our results suggest that the combined effect of strain and doping is responsible for achieving EuTiO$_3$-based metallic FE-FM multiferroics and may provide a new way for obtaining metallic FE-FM multiferroics in other materials.

## 1. Introduction

---

It is usually considered that ferroelectricity cannot exist in metals because the conducting electrons screen static internal electric fields. However, Anderson and Blount[1] predicted in 1965 that ferroelectricity can appear in a metal provided that the electron states at the Fermi level are not coupled to the ferroelectric (FE) distortions. In 2013, Shi et al.[2] found the first solid FE metal, $LiOsO_3$, which transforms from a centrosymmetric phase to a ferroelectric-like phase when cooled down to ~140 K while remains metallic. Since then, a number of studies[3–10] have confirmed the existence of FE metals, also known as polar metals, although the types of material systems exhibiting such properties are scarce. These polar metals have potential applications in non-linear optics[11] and ferroelectric devices,[12] and as topological materials for quantum devices.[13] Meanwhile, ferroelectric-ferromagnetic (FE-FM) multiferroics display simultaneous ferroelectricity and ferromagnetism.[14–17] These materials can enable the control of electric polarization by a magnetic field or the control of magnetic moments by an external electric field,[16] thus can bring about new technologies for low-power and high-density information storage. However, FE-FM multiferroic materials are rare. Searching for new FE-FM multiferroic materials is currently an active research field in material research.

It is exceedingly scarce that ferroelectricity, ferromagnetism, metallicity can coexist simultaneously in a material. So far, the coexistence of these three properties has been demonstrated in superlattices, heterostructures, and layered multidomain structures. Cao et al. created[18] a non-centrosymmetric two-dimensional polar metal of the B-site type in $BaTiO_3/SrTiO_3/LaTiO_3$ superlattices, in which FE, ferromagnetic (FM), and superconducting phases coexist. Meng et al.[19] reported that an FE-FM metal phase is established in $BaTiO_3/SrRuO_3/BaTiO_3$ heterostructures. Using the hybrid density functional theories, Shimada[20] et al. show that metallic conductivity can coexist with FE distortion and layer-arranged ferromagnetism in multidomain electron-doped $PbTiO_3$. It remains an open question whether metallicity, ferroelectricity, and ferromagnetism can coexist within a single domain of a material. For achieving such metallic FE-FM multiferroics, we note that strain is an effective way to induce ferroelectricity,[21–23] while metallicity and ferromagnetism can coexist in materials that feature the Ruderman-Kittel-Kasuya-Yosida (RKKY) mechanism, in which the FM interaction is mediated by the conducting electrons. Therefore, stained RKKY-type materials, such as H-doped $EuTiO_3$,[24] are promising systems for realizing metallic FE-FM multiferroics.

$EuTiO_3$ has been actively studied for potential applications in magnetic sensing and memories,[17] magneto-optical devices,[25] and magnetocaloric energy conversion devices.[26–28] It exhibits a wide range of interesting properties such as multiferroics,[17,29] ferromagnetism,[24,30–33] anomalous Hall effect,[34] two-dimensional electron gas,[35,36] and superconductivity[35], giant magnetocaloric effect.[26–28] The rich behaviors of $EuTiO_3$ come in large part because it stabilizes at a multi-critical equilibrium between paraelectric (PE) and FE states and antiferromagnetic (AFM) and FM states. Without strain or doping,

EuTiO$_3$ shows a PE-AFM state.[37,38] With strain, however, its transitions into the FE-FM state.[17,29,39] The magnetic configuration of EuTiO$_3$ is G-type AFM below 5.3 K or 5.5 K,[40,41] which is delicately balanced between the AFM and FM states.[42,43] On the other hand, doping can not only introduce metallicity into the material but also can alter the balance between AFM and FM states and induce ferromagnetism.[24,30–33] For example, hydride substitution of EuTiO$_3$ results in metallic EuTiO$_{3-x}$H$_x$ accompanied by an AFM-to-FM transition[24], in which the ferromagnetism is achieved by the RKKY interaction between the Eu$^{2+}$ ions mediated via the itinerant Ti 3d electrons. With these experimental observations, strained EuTiO$_{3-x}$H$_x$ film is a promising candidate for metallic FE-FM multiferroics.

In this paper, we calculate the structural, electric, magnetic, and polarization properties of EuTiO$_{3-x}$H$_x$ bulk and films using hybrid density functional theory (DFT) with the PBE0 functional.[44,45] The results demonstrate that bulk EuTiO$_{3-x}$H$_x$ is an AFM insulator at x = 0 and a FM metal at x = 0.125 and 0.25, which agree with experiments. A metal FE FM phase is predicted in strained EuTiO$_{3-x}$H$_x$ films. The appearance of magnetism in polar metals enables using the magnetic field to manipulate these materials in applications. The underlying mechanism for the coexistence of ferroelectricity, ferromagnetism, and metallicity in EuTiO$_{3-x}$H$_x$ films is discussed. Our results suggest that combining strain engineering and doping is a promising way to achieve a EuTiO$_3$-based metallic FE-FM material, and may provide a new way for obtaining other potential metallic FE-FM materials.

## 2. Computational methods

We performed hybrid DFT calculations using a tuned PBE0 hybrid functional that includes 22%[46,47] Hartree-Fock exchange. The amount of exchange is chosen to reproduce a G-AFM ground state of undoped EuTiO$_3$ with a 0.87 eV energy gap as in the experiments.[41] It also predicts a lattice constant of 3.882 Å for cubic EuTiO$_3$ bulk, which agrees with the experimental value of 3.904 Å.[48] We use the projector-augmented wave (PAW) pseudopotentials,[49] in which the valence states include 4f5s5p6s, 3d4s, 2s2p, and 1s electrons for Eu, Ti, O, and H, respectively. We used a 2×2×2 Monkhorst-Pack k-point grid[50] to optimize 40-atom Eu$_8$Ti$_8$O$_{24-x}$H$_x$ (x = 0, 1, 2) supercell and a 3×3×3 Gamma-centered k-point grid to calculate the density of states (DOS) and to get the band gap. The atomic positions of supercells are optimized iteratively until the total-energy difference is 10$^{-3}$ eV or less between two successive ionic steps. The electronic self-consistent iterations are converged to 10$^{-4}$ eV or less between successive iterations using the plane-wave cutoff of 400eV. The calculations are carried out using the Vienna Ab initio Simulation Package (VASP).[51]

EuTiO$_3$ has the cubic structure (space group Pm-3m) with the experimental lattice constants of 3.904 Å.[48] Meanwhile, unstrained hydrogen-doped EuTiO$_{3-x}$H$_x$ stabilizes the cubic perovskite and shows metallic ferromagnetism.[24] In order to compare with experimental electric and magnetic properties of unstrained EuTiO$_{3-x}$H$_x$, in this study, we first build a 2×2×2 supercell of Eu$_8$Ti$_8$O$_{24-x}$H$_x$ (x = 0, 1, 2) with 40 atoms, representing EuTiO$_{3-x}$H$_x$ (x = 0, 0.125, 0.25), and relax both the atomic positions and the unit cell. The four magnetic structures of A-AFM, C-AFM, G-AFM, and FM are considered for the spin configuration.[52] Then we apply the in-plane compressive strain to model the strained EuTiO$_{3-x}$H$_x$ (x = 0.125, 0.25) films that are epitaxially grown along [001] direction on the substrate. The atomic positions and lattice constant along the (001) direction are fully optimized. The FM and G-AFM magnetic structures are considered for the spin-configurations of the strained EuTiO$_{3-x}$H$_x$, because the experiments have shown a G-AFM ground state of undoped EuTiO$_3$ and an FM of hydrated EuTiO$_{3-x}$H$_x$ and our calculated results show that the energy difference between FM and G-AFM states in EuTiO$_3$ while the energies of A-AFM and C-AFM states are much higher than both the FM and G-AFM states.

## 3. Results and discussions

### 3.1 Effects of doping on unstrained EuTiO$_{3-x}$H$_x$

Table 1 shows the lattice constants, band gaps, and magnetic ground states of bulk, unstrained EuTiO$_{3-x}$H$_x$ from the experimental literature and this work. First, we compare crystal structures of bulk EuTiO$_{3-x}$H$_x$ obtained from the calculations with experiments. In our calculations, all the structures of bulk EuTiO$_{3-x}$H$_x$ stabilize with cubic symmetry, which agrees with experiments.[24] According to the present calculations, the value of lattice constant in EuTiO$_3$ is 3.882 Å, which is 0.56% smaller than the experimental value of 3.904 Å.[24] With H content increasing, the values of both experimental and calculated lattice constant increase monotonously. The theoretical lattice constants are slightly smaller than the experimental values, which is understandable as hybrid DFT calculations simulate structures at 0 k, not including thermal expiation. Overall, the calculated lattice parameters and their trend with increasing x in EuTiO$_{3-x}$H$_x$ are in excellent agreement with experiments.[24]

Next, we examine the electronic structures of EuTiO$_{3-x}$H$_x$ (x = 0, 0.125, 0.25). As shown in Table 1, our calculations show that the ground state of EuTiO$_{3-x}$H$_x$ (x = 0, 0.125, 0.25) is a G-AFM insulator at x = 0 and is a FM metal at x = 0.125 and 0.25, which agrees with available experiments.[24] The transition from an insulator to a metal upon hydrogen doping can be seen from the partial DOS (PDOS) of Eu 4f, Ti 3d, O 2p, H 1s bands of EuTiO$_{3-x}$H$_x$ shown in Figure 1. The energy window under our considerations is from −10 eV to 5 eV. The Fermi level is at 0 eV.

Table 1. Experimental and theoretical lattice constants, band gaps, and magnetic ground states of bulk $EuTiO_{3-x}H_x$.

| x | | a (Å) | Band gap (eV) | The magnetic ground state |
|---|---|---|---|---|
| 0 | This work | 3.882 | 0.87 | G-AFM |
| | Experiment | 3.904[48] | 0.80/0.93[41,48] | G-AFM[48,53] |
| 0.07 | Experiment[24] | 3.906 | 0 | FM |
| 0.125 | This work | 3.885 | 0 | FM |
| 0.15 | Experiment[24] | 3.909 | 0 | FM |
| 0.25 | This work | 3.887 | 0 | FM |
| 0.3 | Experiment[24] | 3.914 | 0 | FM |

Experiments have shown that undoped $EuTiO_3$ is an AFM insulator[48] with an optical band gap of $0.93 \pm 0.07$ eV or 0.8 eV,[41,48] which are in agreement with the present calculated value of 0.87 eV. Figure 1(a) shows that the PDOS of Ti 3d and O 2p are symmetric as the ground state of $EuTiO_3$ is G-AFM. The occupied bands below the Fermi level are composed of Eu 4f, Ti 3d and O 2p bands and hybridization appears between them. The narrow Eu 4f bands below the Fermi level suggest these states are localized. The bottom of the conduction bands consists of Ti 3d and Eu 4f states. According to our calculations, an insulating band gap of 0.87 eV opens between the valence band top and the conduction band bottom at Γ point, which agrees with the experimental[41,48] and previous computational studies using the hybrid density functional approach.[43,48]

For hydrated $EuTiO_{3-x}H_x$ (x = 0.125, 0.25), the PDOS (Figs. 1(b) and 1(c)) are different from those of undoped $EuTiO_3$. The PDOS of Ti 3d and O 2p are not symmetric, which indicates spin polarization of Ti states and the occurrence of the FM ground state. Meanwhile, metallic characteristics appear in $EuTiO_{2.875}H_{0.125}$ and $EuTiO_{2.75}H_{0.25}$. The substitutional H donates one itinerant electron to the system and introduces the appearance of $Ti^{3+}(3d^1)$, as shown by the appearance of new Ti states near the Fermi level in Figs 1(b) and 1(c). As a result, the bands straddling the Fermi level are dominated by the Ti 3d states, resulting in a metallic ground state. The PDOS at the Fermi level is ~0.1 states/eV. Through this mechanism, the H substitution induces both the AFM-to-FM and insulator-to-metal transitions in $EuTiO_{3-x}H_x$. Experimentally, the resistivity of $EuTiO_{3-x}H_x$[24] shows metallic characteristics, and an AFM-to-FM transition is observed at x = ~0.07, which verifies the present calculated results. Obvious hybridization appears between Eu 4f, Ti 3d, O 2p, and H 1s states below the Fermi level. Hybridization between Ti 3d and Eu 4f bands of $EuTiO_{3-x}H_x$ (x = 0.125, 0.25) near the Fermi level is stronger than that of $EuTiO_3$, as indicated by the broadening of these states. This result suggests that the FM interaction of itinerant Ti 3d electrons with the localized Eu 4f electron becomes stronger by the d-f exchange.

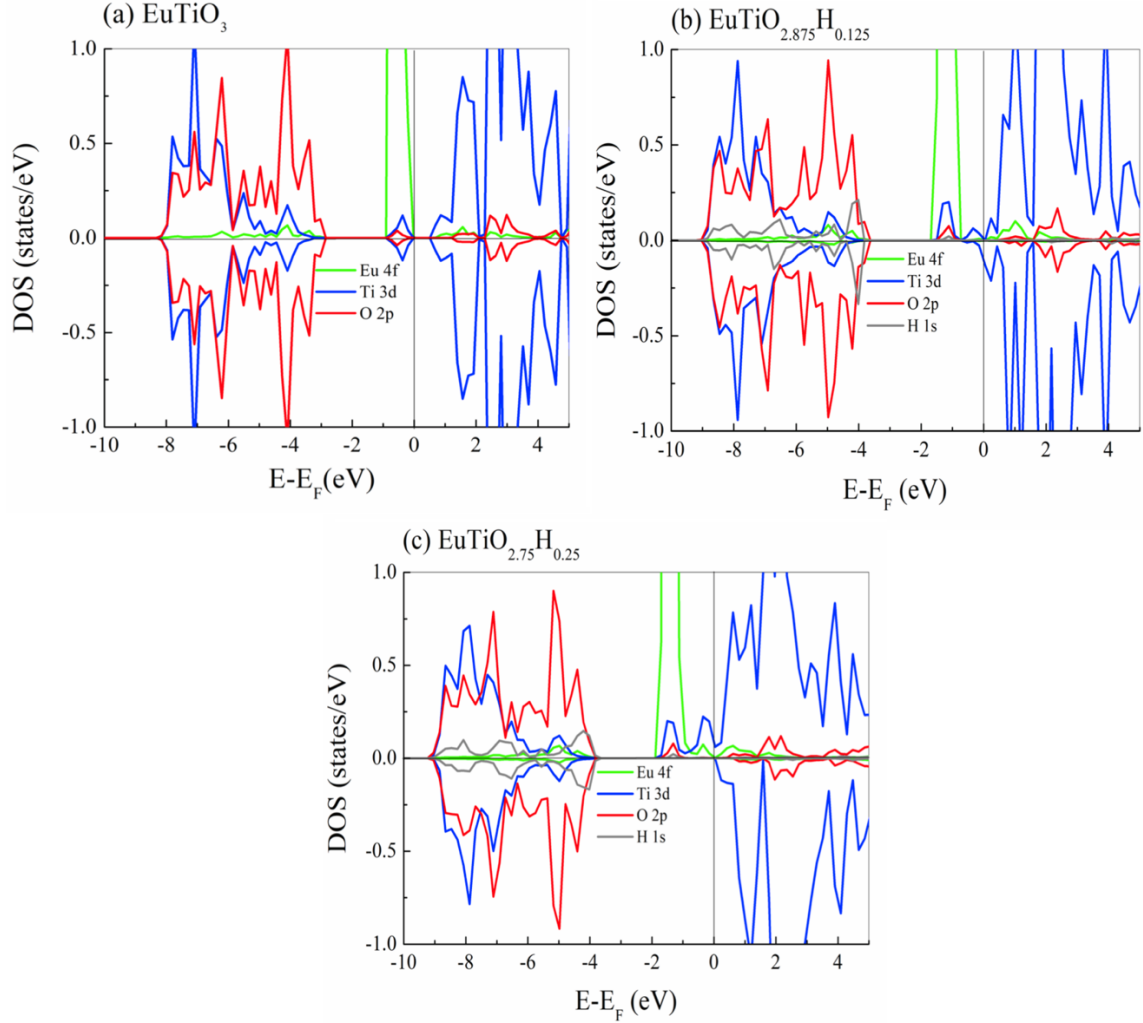

Figure 1. PDOS of bulk EuTiO$_{3-x}$H$_x$: (a) x = 0, (b) x = 0.125, (c) x = 0.25.

In order to understand the experimentally-observed ferromagnetism,[24] we calculated the total energies per formal unit (f.u.) of the FM and A-, C-, G-AFM states, the magnetic moments of Eu atoms, and total magnetic moments per f.u. of unstrained EuTiO$_{3-x}$H$_x$ (x = 0, 0.125, 0.25). The calculated results show that the ground state is G-AFM at x = 0 and FM at x = 0.125, 0.25 in bulk EuTiO$_{3-x}$H$_x$, which agrees with experiments.[24,40] Figure 2 shows the total energy difference between the FM and G-AFM states, the magnetic moments of Eu, and total magnetic moments per f.u. of unstrained EuTiO$_{3-x}$H$_x$. For EuTiO$_3$, the total energies of FM, A-AFM, C-AFM states are higher than that of G-AFM state by 0.6 meV, 3.0 meV, and 4.8 meV/f.u., respectively. The very small difference in energy between the G-AFM and FM states suggests that the G-AFM superexchange and FM exchange are delicately balanced in EuTiO$_3$ although the ground state is G-AFM and the total magnetic moment is 0. The substitution of H for O changes the balance between the G-AFM and FM states. As shown in Figure 2, the total energy of FM is lower than that of G-AFM in bulk $EuTiO_{3-x}H_x$ by 3.3 meV/f.u. and 39.1 meV/f.u. at x = 0.125 and 0.25, respectively. That implies that FM interaction becomes much stronger in bulk $EuTiO_{3-x}H_x$ as H content x increases. Therefore, the total magnetic moment and the Eu atom moment increase with increasing x.

The ferromagnetism induced by H substitution in $EuTiO_{3-x}H_x$ can be interpreted by the RKKY interaction between the Eu atoms mediated by the itinerant Ti 3d electrons. Substitution by H for O converts some of the Ti ions from the $Ti^{4+}(3d^0)$ to the $Ti^{3+}(3d^1)$ configuration in $EuTiO_{3-x}H_x$, therefore introducing itinerant electrons to the system. As shown in Figure 1(b) and 1(d), itinerant electrons occupy the bottom of the conduction band, and the metallicity appears. With the appearance of itinerant electrons, ferromagnetism occurs, and at the same time, hybridization between Ti 3d and Eu 4f bands in $EuTiO_{3-x}H_x$ becomes stronger (see Figure 1) than that of $EuTiO_3$. All these observations are consistent with the interpretation that the ferromagnetism in hydrated $EuTiO_{3-x}H_x$ is caused by the RKKY mechanism through the itinerant electrons in the $Ti^{3+}(3d^1)$ states introduced by H substitution. As H content x increases, itinerant Ti electrons increases, and then the FM interaction between $Eu^{2+}$ spins mediated by these itinerant electrons becomes stronger. As a result, the total energy difference between the FM and G-AFM states, the total magnetic moments, and the Eu magnetic moment in unstrained $EuTiO_{3-x}H_x$ increases with increasing x.

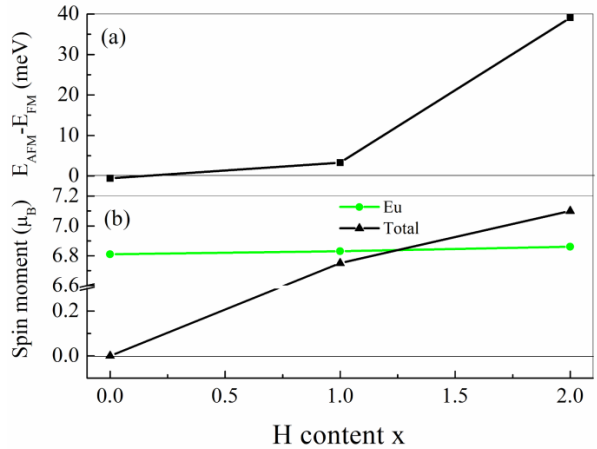

Figure 2. (a) The total energy difference between the FM and G-AFM states per formal unit and (b) the calculated magnetic moment of a Eu atom and total magnetic moments per formal unit of bulk $EuTiO_{3-x}H_x$.

### 3.2 Combined effects of doping and strain on $EuTiO_{3-x}H_x$ films

We now examine the strained $EuTiO_{3-x}H_x$ films that are epitaxially grown along [001] directions. These films have a square-basis lattice of parameter $a_f$ that matches the lattice parameters of the substrates. We define the misfit strain as $\eta = (a_f - a_b)/a_b$, where $a_b$ is the in-plane lattice constant of optimized bulk $EuTiO_{3-x}H_x$. With a certain fixed value of $a_f$, the c-axis and all the atoms in $EuTiO_{3-x}H_x$ films are fully relaxed in the process of structural optimization. Since the FE polarization in metallic materials is not a defined physical quantity, we define a new physical quantity P* to characterize the polar distortion in $EuTiO_{3-x}H_x$ film. $P^* = B_l - B_s$, where $B_l$ and $B_s$ represent the lengths of the long and short Ti-O bonds, respectively, as shown in Figure 3. When $P^* = 0$, there is no polar distortion. Moreover, longer P* represents larger polar distortion. Here we also define a total energy difference per f.u. $\Delta E_{FE-PE}$ with $\Delta E_{FE-PE} = E_{FE} - E_{PE}$, where $E_{FE}$ and $E_{PE}$ represent the total energy per f.u. of FE FM (FE-FM) state and PE FM (PM-FM) state, respectively, and a total energy difference $\Delta E_{FM-AFM}$ with $\Delta E_{FM-AFM} = E_{FM} - E_{AFM}$, where $E_{FM}$ and $E_{AFM}$ represent the total energy per f.u. of FE-FM state and FE AFM (FE-AFM) state, respectively. More negative $\Delta E_{FE-PE}$ means suggests a more stable FM polar structure. Furthermore, more negative $\Delta E_{FM-AFM}$ means a more difficult transition from an FE-FM phase to an FE-AFM phase.

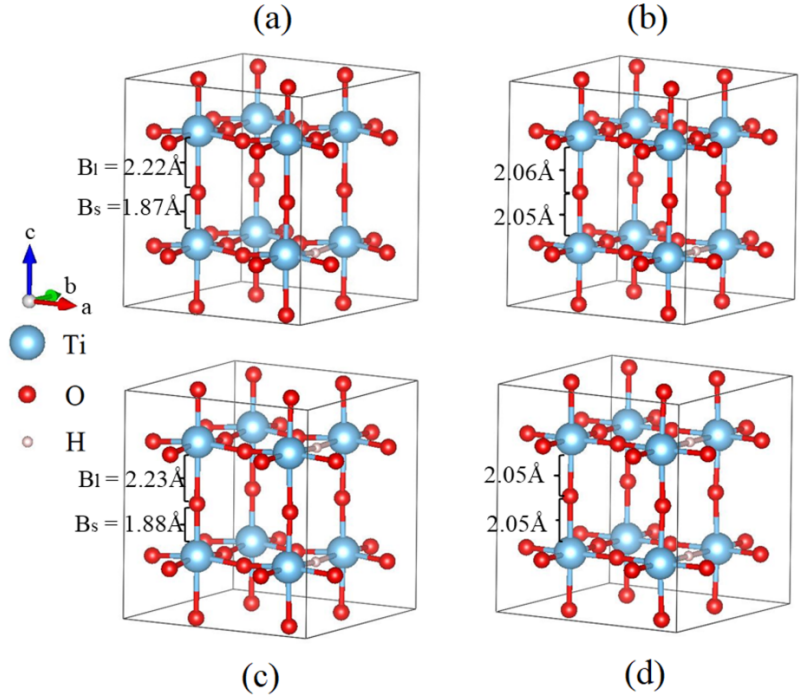

Figure 3. Optimized structures of strained $EuTiO_{3-x}H_x$ at $\eta = -3\%$: (a) FE-FM structure at x = 0.125, (b) PE-FM structure at x = 0.125, (c) FE-FM structure at x = 0.25, (d) PE-FM structure x = 0.25. Because the polarization occurs between Ti-O bonds, we do not show the Eu atoms in the figure.

To study the combined effect of H substitution and compressive strain on the polar distortion and ferromagnetism of EuTiO$_{3-x}$H$_x$, we calculated P*, $\Delta E_{FE-PE}$, and $\Delta E_{FM-AFM}$ of strained EuTiO$_{3-x}$H$_x$ film as functions of the misfit train η in Figure 4. First, EuTiO$_{2.875}$H$_{0.125}$ film is considered. As shown in Fig. 4 (a), the effect of compressive strain on EuTiO$_{2.875}$H$_{0.125}$ film can be divided into two stages: η ≥ −2.4% and η < −2.4%. As aforementioned, the bulk EuTiO$_{2.875}$H$_{0.125}$ (η = 0) shows metallic ferromagnetism, which can be interpreted by RKKY interaction between the Eu mediated by the itinerant Ti 3d electrons. The bulk EuTiO$_{2.875}$H$_{0.125}$ does not show ferroelectricity because the total energy of the FE-FM phase is higher than that of the PE-FM phase. For 0 > η ≥ −2.4%, $\Delta E_{FE-PE}$ in EuTiO$_{2.875}$H$_{0.125}$ film is ~ −0.1 meV/f.u. with P* < 0.06 Å, which implies that the strength of FE polarization is exceedingly weak. $\Delta E_{FM-AFE}$ decreases slightly with the increase of compressive strain at 0 > η ≥ −2.4%. Therefore, for η ≥ −2.4%, EuTiO$_{2.875}$H$_{0.125}$ thin film shows mainly PE metallic ferromagnetism, where the H-doping is responsible for both the metallicity and the ferromagnetism through the RKKY mechanism.

As the compressive strain increases further, the situation changes. The energy difference between the FM and AFM phases decreases slowly first and then stabilizes at ~ −24.0 meV/f.u.. $\Delta E_{FM-AFE}$ is −28.7 meV/f.u. and −24.0 meV/f.u. at η = −2.4% and η = −2.7%, respectively. However, the energy difference between the FE and PE phases increases obviously with the increase of compressive strain at η < −2.4%. $\Delta E_{FE-PE}$ is −11.6 meV/f.u. and −15.1 meV/f.u. at η = −2.7% and −3.0%, respectively. Furthermore, a large polar distortion P*, which is 0.34 Å at η = −2.7%, −3.0%, occurs starting at η < −2.4%. Figure 3 (a) and (b) show the crystal structures of stable FE and PE EuTiO$_{2.875}$H$_{0.125}$ film at η = −3%. These mean that EuTiO$_{2.875}$H$_{0.125}$ film shows a stable FE ferromagnet at η < −2.4%. It can be seen that strong polarization occurs due to large compressive strain. Figure 5 (a) and (b) shows the partial DOS of EuTiO$_{2.875}$H$_{0.125}$ film at η = −2.7%, −3%. Similar to the bulk EuTiO$_{2.875}$H$_{0.125}$, the itinerant Ti 3d states across the Fermi surface lead to the metallic characteristic, which proves that the metallicity of EuTiO$_{2.875}$H$_{0.125}$ film still comes from the H-doping-induced itinerant Ti 3d electrons. Therefore, metallicity, ferroelectricity, and ferromagnetism coexist in EuTiO$_{2.875}$H$_{0.125}$ at η < −2.4%.

After experimentally demonstrating the coexistence of metallicity and ferromagnetism in bulk EuTiO$_{3-x}$H$_x$, Yamamoto et al.[24] suggested that the strained EuTiO$_{3-x}$H$_x$ film is a possible candidate of an FM and FE metal. Our theoretical calculations here demonstrate the hypothesis. Ferroelectricity is usually not considered to exist in metals since conducting electrons screen the static internal electric fields. After LiOsO$_3$, the first solid FE metal was found, a few other polar metals are experimentally (theoretically) observed (predicted).[4,10,18,20,54,55] However, the coexistence of ferroelectricity, ferromagnetism, and metallicity is exceedingly rare, especially within one material. Cao et al.[18] realized room temperature polarized metal in the three-color BaTiO$_3$/SrTiO$_3$/LaTiO$_3$ superlattice, in which ferroelectricity, ferromagnetism, and superconductivity coexist. Meng et al.[19] established an FM polar metal phase is in $BaTiO_3/SrRuO_3/BaTiO_3$ heterostructures. Shimada et al.[20] predicted the coexistence of metallic conductivity, FE distortion and layer-arranged ferromagnetism in multidomain electron-doped $PbTiO_3$. In this work, the coexistence of metallicity, ferroelectricity, and ferromagnetism are realized within a single domain in strained $EuTiO_{3-x}H_x$ thin film, which suggests that combined action of strain engineering and doping is a promising way to achieve a $EuTiO_3$-based metallic FE-FM material. It will provide a new way of obtaining other potential metallic FE-FM materials. The fact that ferromagnetism can exist in ferroelectric metals suggests that it is possible to use the magnetic field to switch the polarization and electronic states of these materials, which gives additional functionally and controllability for device designs in applications.

Next, we focus on $EuTiO_{2.75}H_{0.25}$ film. Similar to the $EuTiO_{2.875}H_{0.125}$ film, the $EuTiO_{2.75}H_{0.25}$ film shows mainly metallic PE ferromagnetism when the strain $\eta \geq -2.7\%$. With the increase of the compressive strain, the spin-lattice coupling is strengthened. When the strain $\eta$ is less than $-2.7\%$, the metallic FE ferromagnetism appears in this system, as shown in Figure 3(c) and Figure 4(b). However, there are three differences in $EuTiO_{2.75}H_{0.25}$ from $EuTiO_{2.875}H_{0.125}$ as follows. First, although the total energy of FM state in bulk $EuTiO_{2.75}H_{0.25}$ ($\eta = 0$) is lower than that of AFM state, the energy difference between the FM and AFM states in $EuTiO_{2.75}H_{0.25}$ is much bigger than that of bulk $EuTiO_{2.875}H_{0.125}$ due to increase of hydrogen content. This result shows that the main FM interaction, which is the RKKY mechanism induced by H doping, is strengthened with the increase of H content. Secondly, a PE-FE phase transition point of $\eta = -2.7\%$ in $EuTiO_{2.75}H_{0.25}$ film, is more negative than that of $\eta = -2.4\%$ in $EuTiO_{2.875}H_{0.125}$ film. Compared with $EuTiO_{3-x}H_x$ films, a less negative value of the transition point at $\eta = -1.1\%$ appears in undoped $EuTiO_3$ thin film,[17] which implies that H doping has a negative effect on the polarization of $EuTiO_{3-x}H_x$ films. Finally, although the polarization distortion of $EuTiO_{2.75}H_{0.25}$ film at $\eta = -3.0\%$ is almost the same as that of $EuTiO_{2.875}H_{0.125}$ film, the total energy difference $\Delta E_{FE-PE}$ between FE-FM and PE-FM states of $EuTiO_{2.75}H_{0.25}$ film is much smaller than that of $EuTiO_{2.875}H_{0.125}$, as shown in Figure 3 and 4. From these three observations, we can conclude that the main contribution of H doping in $EuTiO_{3-x}H_x$ film is to induce metallic characteristics and enhance the RKKY FM interaction, but it is not beneficial for the stability of the electrical polarization. To achieve metallic FE-FM multiferroics, the effect of strain and H doping must be delicately balanced.

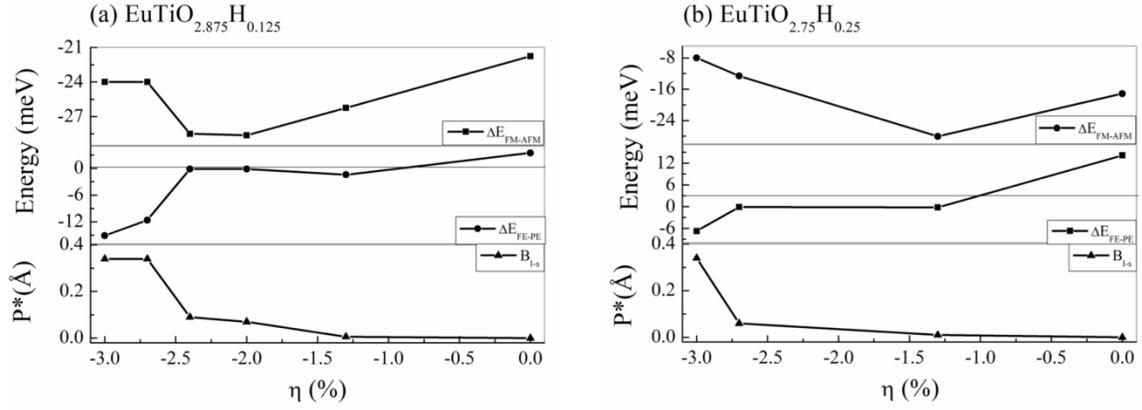

Figure 4. The defined physical parameter P* and the total energy difference per f.u. of strained EuTiO$_{3-x}$H$_x$ at $0 \leq \eta \leq -3\%$: (a) x = 0.125, (b) x = 0.25. In the top figure, $\Delta E_{FM-AFM}$ is defined as the total energy difference per f.u. between the FE-FM and FE-AFM states. In the middle figure, $\Delta E_{FE-PE}$ is defined as the total energy difference per f.u. between the FE-FM and PM-FM states. In the bottom figure, P* is defined as the difference between the lengths of the long and short Ti-O bonds to characterize the polar distortion.

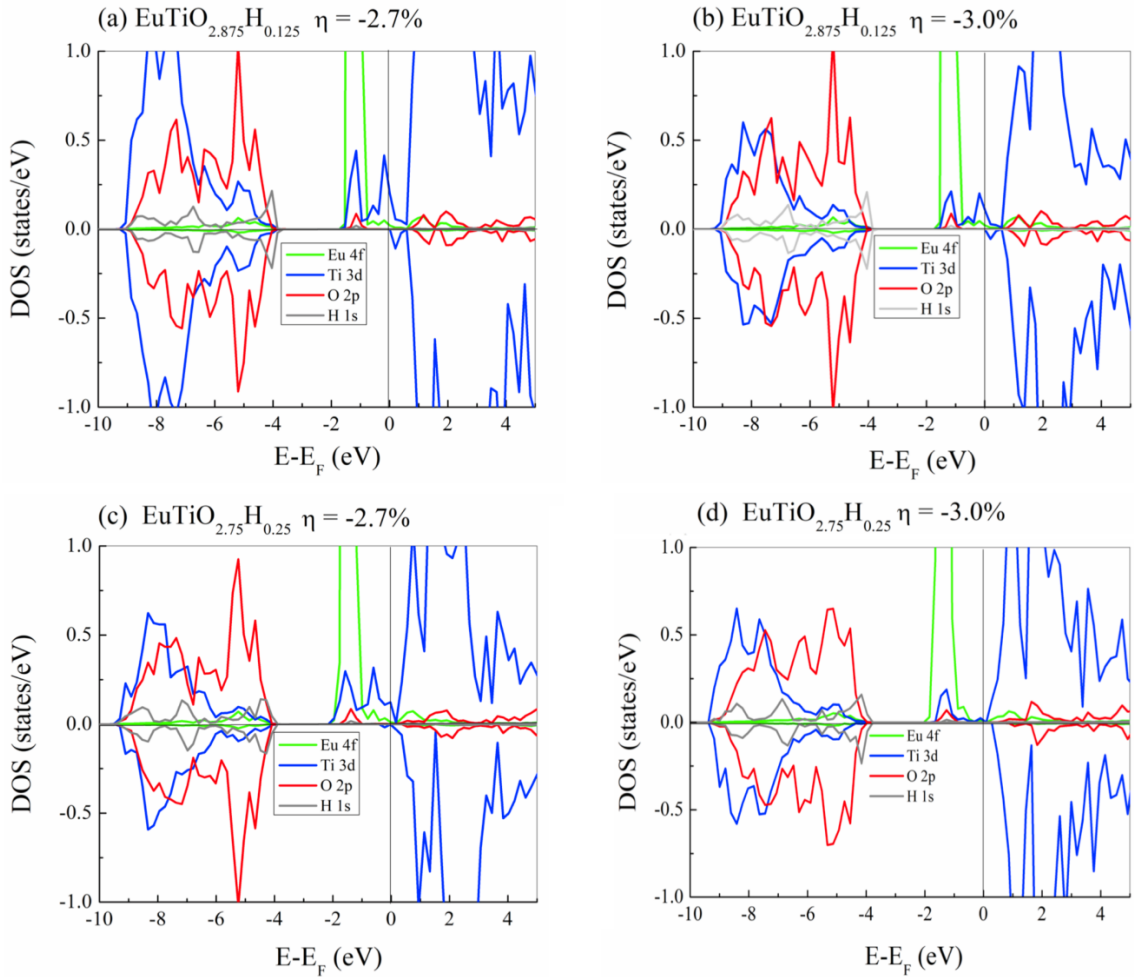

Figure 5. PDOS of FE-FE state in strained EuTiO$_{3-x}$H$_x$: (a) $x = 0.125$, $\eta = -2.7\%$ (b) $x = 0.125$, $\eta = -3.0\%$ (c) $x = 0.25$, $\eta = -2.7\%$ (d) $x = 0.25$, $\eta = -3.0\%$.

Finally, we discuss the mechanism of the coexistence of metallicity, ferroelectricity, and ferromagnetism in EuTiO$_{3-x}$H$_x$ film. EuTiO$_3$ shows G-type antiferromagnetism[40,41] below 5.3K or 5.5K and quantum PE behavior.[38] Experimentally, ferromagnetism is reported in La, Nb, H, and Cr-doped EuTiO$_3$[24,56–58] and an epitaxial EuTiO$_3$ film with an elongated c-axis.[59] Moreover, an FM and FE phase has been realized in an epitaxial film on a DyScO$_3$ substrate using 1.1% biaxial tensile strain.[17] All these facts demonstrate that EuTiO$_3$, which has a strong spin-lattice coupling[17], is multi-critically balanced between the PE and FE states and between the AFM and FM states. In bulk EuTiO$_{3-x}$H$_x$, one H atom contributes an itinerant electron to the system, resulting in the emergence of Ti$^{3+}$ (3d$^1$). As shown in Figs. 1(b) and 1(c), the new Ti 3d state across the Fermi surface leads to the appearance of metallicity. Hybridization between the itinerant Ti 3d and Eu 4f states become stronger with the increase of H doping. Furthermore, the total magnetic moment of the system increase significantly, and the energy difference between the FM and AFM states becomes more negative, indicating that H doping leads to a more stable FM phase, as shown in Figure 2. The ferromagnetism in hydrated EuTiO$_{3-x}$H$_x$ is caused by the RKKY interaction between Eu$^{2+}$ spin through the H-doping-deduced itinerant Ti$^{3+}$(3d$^1$) electrons.

The appearance of an FE-FM phase in EuTiO$_{3-x}$H$_x$ film under compressive strain is due to strong the spin-lattice coupling similar to the undoped EuTiO$_3$.[17] In such systems, the spin-lattice coupling mechanism is described by the formula $\omega^2 = \omega_0^2 - \lambda \langle S_i \cdot S_j \rangle$[17], where $\omega$ is the frequency of a phonon mode that is infrared-active, and $\omega_0$ is the bare phonon frequency without spin-lattice interaction. The macroscopic spin-phonon coupling is denoted by the constant $\lambda$, and $\langle S_i \cdot S_j \rangle$ is the spin–spin correlation function between the nearest-neighbor sites. The application of compressive strain increases the lattice-spin coupling. A FE-FM phase begins to appear at $\eta = -2.4\%$ in EuTiO$_{2.875}$H$_{0.125}$ film and at $\eta = -2.7\%$ in EuTiO$_{2.75}$H$_{0.25}$ film, as shown in Figure 4. With the further increase of compressive strain, the polarization distortion and the stability of the ferroelectric phase gradually increase.

Compared with the strained undoped EuTiO$_3$ thin films, it is found that the $\eta$ value of the PM-FM phase transition point becomes more negative with the increase of H doping. This result suggests that H doping is not beneficial for the appearance and stability of the ferroelectric phase. This can be explained as follows: there are two kinds of Ti ions in H-doped EuTiO$_{3-x}$H$_x$: Ti$^{3+}$ (d$^1$) and Ti$^{4+}$ (d$^0$). The Ti$^{3+}$ (d$^1$) ion has an itinerant electron and is responsible for the metallicity of the system. The Ti$^{4+}$ (d$^0$) ions, on the other hand, are responsible for ferroelectricity under compressive strain, as in the case of undoped EuTiO$_3$ thin films, where all Ti ions are Ti$^{4+}$ (d$^0$). With an increase of H content, the amount of Ti$^{3+}$ (d$^1$) ions increases, but the amount of Ti$^{4+}$ (d$^0$) ions decreases. As a result, FE polarization is more difficult to occur with the increase of H doping.

With the increase of compressive strain, the metallic FE-FM phase appears in the system at η < −2.4 for $EuTiO_{2.875}H_{0.125}$ film and at η < −2.7% for $EuTiO_{2.75}H_{0.25}$ film, as shown in Figure 3, 4, 5. This result means that ferromagnetism, ferroelectricity, and metallicity coexist in the strained $EuTiO_{3-x}H_x$ film. It is well known that strain can induce ferromagnetism and ferroelectricity in $EuTiO_3$ while it has never been reported that ferroelectricity and metallicity can coexist in $EuTiO_3$-based materials. Ferroelectricity is not usually considered to be in metals because conductor electrons screen static internal electric fields. However, Anderson and Blount[1] predicted that ferroelectricity can appear in a metal provided that the electrons at the Fermi level do not couple the FE distortions. For example, $LiOsO_3$, the first solid FE metal, remains conducting and has an FE ionic structure below 140 K.[2] Laurita et al.[6] provided experimental evidence for the weakly coupled electron mechanism in $LiOsO_3$. In order to understand further why metallicity and ferroelectricity coexist in $EuTiO_{3-x}H_x$ films, we demonstrate the PDOS of $EuTiO_{3-x}H_x$ (x = 0.125, 0.25) at η = −2.7%, −3.0% in Figure 5. Compared with DOS of bulk $EuTiO_{3-x}H_x$ (x = 0.125, 0.25) in Figure 1, the metallicity of $EuTiO_{3-x}H_x$ does not change, although there are subtle changes for DOS at the Fermi level of $EuTiO_{3-x}H_x$ as compressive strain increases. However, Figure 4 shows that the electrical polarization P* is almost 0 at η ≥ −2.4% in $EuTiO_{2.875}H_{0.125}$ and η ≥ −2.7% in $EuTiO_{2.75}H_{0.25}$ and increases rapidly with more compressive strain. Therefore, we can conclude that metallicity in $EuTiO_{3-x}H_x$ films is weakly coupled with polarized distortion. As a result, metallicity can exist in the FE and FM state of $EuTiO_{3-x}H_x$ films.

**4. Conclusion**

In conclusion, structural, electric, magnetic, and polarization properties of bulk and strained $EuTiO_{3-x}H_x$ are investigated by hybrid DFT calculations. Unstrained bulk $EuTiO_{3-x}H_x$ shows insulating antiferromagnetism at x = 0 and metallic ferromagnetism at x = 0.125, 0.25, which agree with experiments. RKKY interaction between $Eu^{2+}$ ions mediated by the itinerant Ti 3d electrons is responsible for ferromagnetism in $EuTiO_{3-x}H_x$. Strained $EuTiO_{3-x}H_x$ films show metallic ferromagnetism at 0 ≤ η ≤ −3% and a metallic FE-FM phase is predicted at η < −2.4% in $EuTiO_{2.875}H_{0.125}$ and at η < −2.7% in $EuTiO_{2.75}H_{0.25}$. We discuss the coexistence mechanism of metallicity, ferroelectricity, and ferromagnetism in $EuTiO_{3-x}H_x$ films. The theoretical results suggest that combined action of strain engineering and doping is a promising way to achieve $EuTiO_3$-based metallic FE-FM multiferroics, and it will provide a new way for obtaining other potential metallic FE-FM multiferroic materials.

**Acknowledgement**

S. Xu and Y. Gu are thankful for the financial support from the Doctoral Research Project of JUST (Nos. JKD120114001) and the Open Project of National Laboratory of Solid State Microstructures at Nanjing University. We are grateful to Jiangsu University of Science and Technology (S. Xu and Y. Gu) and the high-performance computing facility at the University of Memphis (X. Shen.) for the award of CPU hours.

(1)  Anderson, P. W.; Blount, E. I. Symmetry Considerations on Martensitic Transformations: "Ferroelectric" Metals? *Phys. Rev. Lett.* **1965**, *14* (7), 217–219.

(2)  Shi, Y.; Guo, Y.; Wang, X.; Princep, A. J.; Khalyavin, D.; Manuel, P.; Michiue, Y.; Sato, A.; Tsuda, K.; Yu, S.; et al. A Ferroelectric-like Structural Transition in a Metal. *Nat. Mater.* **2013**, *12* (11), 1024–1027.

(3)  Fei, Z.; Zhao, W.; Palomaki, T. A.; Sun, B.; Miller, M. K.; Zhao, Z.; Yan, J.; Xu, X.; Cobden, D. H. Ferroelectric Switching of a Two-Dimensional Metal. *Nature* **2018**, *560* (7718), 336–339.

(4)  Kim, T. H.; Puggioni, D.; Yuan, Y.; Xie, L.; Zhou, H.; Campbell, N.; Ryan, P. J.; Choi, Y.; Kim, J.-W.; Patzner, J. R.; et al. Polar Metals by Geometric Design. *Nature* **2016**, *533* (7601), 68–72.

(5)  Lu, J.; Chen, G.; Luo, W.; Íñiguez, J.; Bellaiche, L.; Xiang, H. Ferroelectricity with Asymmetric Hysteresis in Metallic LiOsO3 Ultrathin Films. *Phys. Rev. Lett.* **2019**, *122* (22), 227601.

(6)  Laurita, N. J.; Ron, A.; Shan, J.-Y.; Puggioni, D.; Koocher, N. Z.; Yamaura, K.; Shi, Y.; Rondinelli, J. M.; Hsieh, D. Evidence for the Weakly Coupled Electron Mechanism in an Anderson-Blount Polar Metal. *Nat. Commun.* **2019**, *10* (1), 3217.

(7)  Rischau, C. W.; Lin, X.; Grams, C. P.; Finck, D.; Harms, S.; Engelmayer, J.; Lorenz, T.; Gallais, Y.; Fauqué, B.; Hemberger, J.; et al. A Ferroelectric Quantum Phase Transition inside the Superconducting Dome of Sr1−xCaxTiO3−δ. *Nat. Phys.* **2017**, *13* (7), 643–648.

(8)  Kamitani, M.; Bahramy, M. S.; Nakajima, T.; Terakura, C.; Hashizume, D.; Arima, T.; Tokura, Y. Superconductivity at the Polar-Nonpolar Phase Boundary of SnP with an Unusual Valence State. *Phys. Rev. Lett.* **2017**, *119* (20), 207001.

(9)  Ghosh, S.; Borisevich, A. Y.; Pantelides, S. T. Engineering an Insulating Ferroelectric Superlattice with a Tunable Band Gap from Metallic Components. *Phys. Rev. Lett.* **2017**, *119* (17), 177603.

(10) Filippetti, A.; Fiorentini, V.; Ricci, F.; Delugas, P.; Íñiguez, J. Prediction of a Native Ferroelectric Metal. *Nat. Commun.* **2016**, *7* (1), 11211.

(11) Padmanabhan, H.; Park, Y.; Puggioni, D.; Yuan, Y.; Cao, Y.; Gasparov, L.; Shi, Y.; Chakhalian, J.; Rondinelli, J. M.; Gopalan, V. Linear and Nonlinear Optical Probe of the Ferroelectric-like Phase Transition in a Polar Metal, LiOsO 3. *Appl. Phys. Lett.* **2018**, *113* (12), 122906.

(12) Puggioni, D.; Giovannetti, G.; Rondinelli, J. M. Polar Metals as Electrodes to Suppress the Critical-Thickness Limit in Ferroelectric Nanocapacitors. *J. Appl. Phys.* **2018**, *124* (17), 174102.

(13) Tokura, Y.; Nagaosa, N. Nonreciprocal Responses from Non-Centrosymmetric Quantum Materials. *Nat. Commun.* **2018**, *9* (1), 3740.

(14) Eerenstein, W.; Mathur, N. D.; Scott, J. F. Multiferroic and Magnetoelectric Materials. *Nature* **2006**, *442* (7104), 759–765.

(15) Ramesh, R.; Spaldin, N. A. Multiferroics: Progress and Prospects in Thin Films. *Nat. Mater.* **2007**, *6* (1), 21–29.

(16) Mandal, P.; Pitcher, M. J.; Alaria, J.; Niu, H.; Borisov, P.; Stamenov, P.; Claridge, J. B.; Rosseinsky, M. J. Designing Switchable Polarization and Magnetization at Room Temperature in an Oxide. *Nature* **2015**, *525* (7569), 363–366.

(17) Lee, J. H.; Fang, L.; Vlahos, E.; Ke, X.; Jung, Y. W.; Kourkoutis, L. F.; Kim, J.-W.; Ryan, P. J.; Heeg, T.; Roeckerath, M.; et al. A Strong Ferroelectric Ferromagnet Created by Means of Spin–lattice Coupling. *Nature* **2010**, *466* (7309), 954–958.

(18) Cao, Y.; Wang, Z.; Park, S. Y.; Yuan, Y.; Liu, X.; Nikitin, S. M.; Akamatsu, H.; Kareev, M.; Middey, S.; Meyers, D.; et al. Artificial Two-Dimensional Polar Metal at Room Temperature. *Nat. Commun.* **2018**, *9* (1), 1547.

(19) Meng, M.; Wang, Z.; Fathima, A.; Ghosh, S.; Saghayezhian, M.; Taylor, J.; Jin, R.; Zhu, Y.; Pantelides, S. T.; Zhang, J.; et al. Interface-Induced Magnetic Polar Metal Phase in Complex Oxides. *Nat. Commun.* **2019**, *10* (1), 5248.

(20) Shimada, T.; Xu, T.; Araki, Y.; Wang, J.; Kitamura, T. Unusual Metallic Multiferroic Transitions in Electron-Doped PbTiO 3. *Adv. Electron. Mater.* **2017**, *3* (8), 1700134.

(21) Haeni, J. H.; Irvin, P.; Chang, W.; Uecker, R.; Reiche, P.; Li, Y. L.; Choudhury, S.; Tian, W.; Hawley, M. E.; Craigo, B.; et al. Room-Temperature Ferroelectricity in Strained SrTiO3. *Nature* **2004**, *430* (7001), 758–761.

(22) Lee, J. H.; Rabe, K. M. Epitaxial-Strain-Induced Multiferroicity in SrMnO3 from First Principles. *Phys. Rev. Lett.* **2010**, *104* (20), 207204.

(23) Gich, M.; Fina, I.; Morelli, A.; Sánchez, F.; Alexe, M.; Gàzquez, J.; Fontcuberta, J.; Roig, A. Multiferroic Iron Oxide Thin Films at Room Temperature. *Adv. Mater.* **2014**, *26* (27), 4645–4652. https://doi.org/10.1002/adma.201400990.

(24) Yamamoto, T.; Yoshii, R.; Bouilly, G.; Kobayashi, Y.; Fujita, K.; Kususe, Y.; Matsushita, Y.; Tanaka, K.; Kageyama, H. An Antiferro-to-Ferromagnetic Transition in EuTiO3–xHx Induced by Hydride Substitution. *Inorg. Chem.* **2015**, *54* (4), 1501–1507.

(25) Bussmann-Holder, A.; Roleder, K.; Stuhlhofer, B.; Logvenov, G.; Lazar, I.; Soszyński, A.; Koperski, J.; Simon, A.; Köhler, J. Transparent EuTiO3 Films: A Possible Two-Dimensional Magneto-Optical Device. *Sci. Rep.* **2017**, *7* (1), 40621.

(26) Midya, A.; Mandal, P.; Rubi, K.; Chen, R.; Wang, J.-S.; Mahendiran, R.; Lorusso, G.; Evangelisti, M. Large Adiabatic Temperature and Magnetic Entropy Changes in EuTiO3. *Phys. Rev. B* **2016**, *93* (9), 094422.

(27) Rubi, K.; Kumar, P.; Maheswar Repaka, D. V.; Chen, R.; Wang, J.-S.; Mahendiran, R. Giant Magnetocaloric Effect in Magnetoelectric Eu 1-X Ba X TiO 3. *Appl. Phys. Lett.* **2014**, *104* (3), 032407.

(28) Roy, S.; Khan, N.; Mandal, P. Giant Low-Field Magnetocaloric Effect in Single-Crystalline EuTi 0.85 Nb 0.15 O 3. *APL Mater.* **2016**, *4* (2), 026102.

(29) Fennie, C. J.; Rabe, K. M. Magnetic and Electric Phase Control in Epitaxial EuTiO3 from First Principles. *Phys. Rev. Lett.* **2006**, *97* (26), 267602.

(30) Takahashi, K. S.; Onoda, M.; Kawasaki, M.; Nagaosa, N.; Tokura, Y. Control of the Anomalous Hall Effect by Doping in Eu1-xLaxO3 Thin Films. *Phys. Rev. Lett.* **2009**, *103* (5), 057204.

(31) Li, L.; Zhou, H.; Yan, J.; Mandrus, D.; Keppens, V. Research Update: Magnetic Phase Diagram of EuTi 1−x B X O 3 ( B = Zr, Nb). *APL Mater.* **2014**, *2* (11), 110701.

(32) Mo, Z.-J.; Sun, Q.-L.; Han, S.; Zhao, Y.; Chen, X.; Li, L.; Liu, G.-D.; Meng, F.-B.; Shen, J. Effects of Mn-Doping on the Giant Magnetocaloric Effect of EuTiO3 Compound. *J. Magn. Magn. Mater.* **2018**, *456*, 31–37.

(33) Mo, Z.; Sun, Q.; Shen, J.; Wang, C.; Meng, F.; Zhang, M.; Huo, Y.; Li, L.; Liu, G. A Giant Magnetocaloric Effect in EuTi 0.875 Mn 0.125 O 3 Compound. *J. Alloys Compd.* **2018**, *753*, 1–5.

(34) Takahashi, K. S.; Ishizuka, H.; Murata, T.; Wang, Q. Y.; Tokura, Y.; Nagaosa, N.; Kawasaki, M. Anomalous Hall Effect Derived from Multiple Weyl Nodes in High-Mobility EuTiO 3 Films. *Sci. Adv.* **2018**, *4* (7), eaar7880.

(35) Stornaiuolo, D.; Cantoni, C.; De Luca, G. M.; Di Capua, R.; Di Gennaro, E.; Ghiringhelli, G.; Jouault, B.; Marrè, D.; Massarotti, D.; Miletto Granozio, F.; et al. Tunable Spin Polarization and Superconductivity in Engineered Oxide Interfaces. *Nat. Mater.* **2016**, *15* (3), 278–283.

(36) Gui, Z.; Janotti, A. Carrier-Density-Induced Ferromagnetism in EuTiO3 Bulk and Heterostructures. *Phys. Rev. Lett.* **2019**, *123* (12), 127201.

(37) Bessas, D.; Rushchanskii, K. Z.; Kachlik, M.; Disch, S.; Gourdon, O.; Bednarcik, J.; Maca, K.; Sergueev, I.; Kamba, S.; Ležaić, M.; et al. Lattice Instabilities in Bulk EuTiO3. *Phys. Rev. B* **2013**, *88* (14), 144308.

(38) Katsufuji, T.; Takagi, H. Coupling between Magnetism and Dielectric Properties in Quantum Paraelectric EuTiO3. *Phys. Rev. B* **2001**, *64* (5), 054415.

(39) Yang, Y.; Ren, W.; Wang, D.; Bellaiche, L. Understanding and Revisiting Properties of $EuTiO_3$ Bulk Material and Films from First Principles. *Phys. Rev. Lett.* **2012**, *109* (26), 267602.

(40) McGuire, T. R.; Shafer, M. W.; Joenk, R. J.; Alperin, H. A.; Pickart, S. J. Magnetic Structure of EuTiO 3. *J. Appl. Phys.* **1966**, *37* (3), 981–982.

(41) Lee, J. H.; Ke, X.; Podraza, N. J.; Kourkoutis, L. F.; Heeg, T.; Roeckerath, M.; Freeland, J. W.; Fennie, C. J.; Schubert, J.; Muller, D. A.; et al. Optical Band Gap and Magnetic Properties of Unstrained EuTiO3 Films. *Appl. Phys. Lett.* **2009**, *94* (21), 212509.

(42) Ranjan, R.; Sadat Nabi, H.; Pentcheva, R. Electronic Structure and Magnetism of EuTiO 3 : A First-Principles Study. *J. Phys. Condens. Matter* **2007**, *19* (40), 406217.

(43) Akamatsu, H.; Kumagai, Y.; Oba, F.; Fujita, K.; Murakami, H.; Tanaka, K.; Tanaka, I. Antiferromagnetic Superexchange via 3d States of Titanium in EuTiO3 as Seen from Hybrid Hartree-Fock Density Functional Calculations. *Phys. Rev. B* **2011**, *83* (21), 214421.

(44) Adamo, C.; Barone, V. Toward Reliable Density Functional Methods without Adjustable Parameters: The PBE0 Model. *J. Chem. Phys.* **1999**, *110* (13), 6158.

(45) Perdew, J. P.; Ernzerhof, M.; Burke, K. Rationale for Mixing Exact Exchange with Density Functional Approximations. *J. Chem. Phys.* **1996**, *105*, 9982.

(46) Xu, S.; Gu, Y.; Wu, X. Structural, Electronic and Magnetic Properties of a Ferromagnetic

Metal: Nb-Doped EuTiO3. *J. Magn. Magn. Mater.* **2020**, *497*, 166077.

(47) Xu, S.; Shen, X.; Hallman, K. A.; Haglund, R. F.; Pantelides, S. T. Unified Band-Theoretic Description of Structural, Electronic, and Magnetic Properties of Vanadium Dioxide Phases. *Phys. Rev. B* **2017**, *95* (12), 125105.

(48) Akamatsu, H.; Fujita, K.; Hayashi, H.; Kawamoto, T.; Kumagai, Y.; Zong, Y.; Iwata, K.; Oba, F.; Tanaka, I.; Tanaka, K. Crystal and Electronic Structure and Magnetic Properties of Divalent Europium Perovskite Oxides EuMO3 ( M = Ti, Zr, and Hf): Experimental and First-Principles Approaches. *Inorg. Chem.* **2012**, *51* (8), 4560–4567.

(49) Kresse, G.; Joubert, D. From Ultrasoft Pseudopotentials to the Projector Augmented-Wave Method. *Phys. Rev. B* **1999**, *59* (3), 1758–1775.

(50) Monkhorst, H. J.; Pack, J. D. Special Points for Brillouin-Zone Integrations. *Phys. Rev. B* **1976**, *13* (12), 5188–5192.

(51) Kresse, G.; Furthmüller, J. Efficient Iterative Schemes for Ab Initio Total-Energy Calculations Using a Plane-Wave Basis Set. *Phys. Rev. B* **1996**, *54* (16), 11169–11186.

(52) Wollan, E. O.; Koehler, W. C. Neutron Diffraction Study of the Magnetic Properties of the Series of Perovskite-Type Compounds [(1-x)La, xCa]MnO{3}. *Phys. Rev.* **1955**, *100* (2), 545–563.

(53) Akahoshi, D.; Koshikawa, S.; Nagase, T.; Wada, E.; Nishina, K.; Kajihara, R.; Kuwahara, H.; Saito, T. Magnetic Phase Diagram for the Mixed-Valence Eu Oxide EuTi1−xAlxO3 (0 ≤ X ≤ 1). *Phys. Rev. B* **2017**, *96* (18), 184419.

(54) Puggioni, D.; Rondinelli, J. M. Designing a Robustly Metallic Noncenstrosymmetric Ruthenate Oxide with Large Thermopower Anisotropy. *Nat. Commun.* **2014**, *5* (1), 3432.

(55) Ma, C.; Jin, K.; Ge, C.; Yang, G. Strain-Engineering Stabilization of BaTiO3-Based Polar Metals. *Phys. Rev. B* **2018**, *97* (11), 115103.

(56) Roy, S.; Khan, N.; Mandal, P. Unconventional Transport Properties of the Itinerant Ferromagnet EuTi1−xNbxO3 (x =0.10–0.20). *Phys. Rev. B* **2018**, *98* (13), 134428.

(57) Wei, T.; Song, Q. G.; Zhou, Q. J.; Li, Z. P.; Qi, X. L.; Liu, W. P.; Guo, Y. R.; Liu, J.-M. Cr-Doping Induced Ferromagnetic Behavior in Antiferromagnetic EuTiO3 Nanoparticles. *Appl. Surf. Sci.* **2011**, *258* (1), 599–603.

(58) Rubi, K.; Midya, A.; Mahendiran, R.; Maheswar Repaka, D. V.; Ramanujan, R. V. Magnetocaloric Properties of Eu 1− X La X TiO 3 (0.01 ≤ X ≤ 0.2) for Cryogenic Magnetic Cooling. *J. Appl. Phys.* **2016**, *119* (24), 243901.

(59) Fujita, K.; Wakasugi, N.; Murai, S.; Zong, Y.; Tanaka, K. High-Quality Antiferromagnetic EuTiO3 Epitaxial Thin Films on SrTiO3 Prepared by Pulsed Laser Deposition and Postannealing. *Appl. Phys. Lett.* **2009**, *94* (6), 062512.